\documentclass[twocolumn,twocolappendix]{aastex701}

\usepackage{bm}

\graphicspath{{./}{figures/}}

\begin{document}

\title{NAPTIME: A Neural-Process Framework for Rubin Alert Classification}

\author[orcid=0000-0003-1714-7415,sname='Earl']{Nicholas~Earl}
\affiliation{Department of Astronomy, University of Illinois Urbana-Champaign, 1002 W. Green St., Urbana, IL 61801, USA}
\affiliation{Center for Astrophysical Surveys, National Center for Supercomputing Applications, Urbana, IL, 61801, USA}
\email[show]{nmearl2@illinois.edu}

\author[orcid=0000-0001-5078-5457,gname=Siddharth, sname=Chaini]{Siddharth Chaini} 
\altaffiliation{NASA FINESST Fellow}
\affiliation{Department of Physics and Astronomy, University of Delaware, Newark, DE 19716, USA}
\affiliation{University of Delaware,
Data Science Institute,
Newark, DE 19716, USA}
\email{chaini@udel.edu}

\author[orcid=0000-0002-4235-7337,sname='French']{K.~Decker~French}
\affiliation{Department of Astronomy, University of Illinois Urbana-Champaign, 1002 W. Green St., Urbana, IL 61801, USA}
\affiliation{Center for Astrophysical Surveys, National Center for Supercomputing Applications, Urbana, IL, 61801, USA}
\email{deckerkf@illinois.edu}

\author[0000-0001-9668-2920]{Jason T. Hinkle}
\altaffiliation{NHFP Einstein Fellow}
\affiliation{Department of Astronomy, University of Illinois Urbana-Champaign, 1002 W. Green St., Urbana, IL 61801, USA} 
\affiliation{NSF-Simons AI Institute for the Sky (SkAI), 172 E. Chestnut St., Chicago, IL 60611, USA}
\email{jhinkle6@illinois.edu}

\author[0009-0004-0436-0932]{Yashasvi Moon}
\affiliation{Department of Astronomy, University of Illinois Urbana-Champaign, 1002 W. Green St., Urbana, IL 61801, USA} 
\email{ymoon@illinois.edu}

\author[0009-0005-1158-1896]{Margaret~Shepherd}
\affiliation{Department of Astronomy, University of Illinois Urbana-Champaign, 1002 W. Green St., Urbana, IL 61801, USA}
\email{ms169@illinois.edu}

\begin{abstract}
The Vera C. Rubin Observatory Legacy Survey of Space and Time will produce a high-volume stream of irregularly sampled multiband alerts for which spectroscopic confirmation will be available only for a small minority of sources. Tidal disruption events are rare phenomena that provide a direct probe of dormant massive black holes, but their light curves can be confused with nuclear variability and other transient subclasses. We present NAPTIME (Neural Astrophysical Photometric Transient Identification and Modeling Engine), a neural-process framework for photometric transient classification under sparse and partial observational context. NAPTIME models irregular multiband light curves directly, combining probabilistic light-curve reconstruction with classification and optional host-galaxy context, as well as photometric-redshift information. We evaluate on two simulated benchmarks: ELAsTiCC2, our primary Rubin-like broad-classification benchmark, and MALLORN, a photometry-only TDE-focused benchmark. On the 15-family ELAsTiCC2 task, the metadata-aware model reaches macro $\mathrm{F1} = 0.903$ and macro \added{area under the receiver operating characteristic curve} ($\mathrm{AUROC}$) $= 0.991$, while a matched photometry-only variant reaches 0.874 and 0.986. Viewed as a TDE-versus-rest ranking model, the classifier yields TDE average precision 0.985 with metadata and 0.979 without. Metadata is most valuable in the low-context regime. Using only the earliest 10\% of retained photometric measurements, macro F1 is $\sim$0.42 with metadata and $\sim$0.34 without it. On MALLORN, NAPTIME reaches macro $\mathrm{F1} = 0.693$ and macro $\mathrm{AUROC} = 0.958$. These results show that neural processes provide a practical probabilistic framework for Rubin-like transient classification and remain effective for TDE-focused candidate recovery.
\end{abstract}


\section{Introduction}\label{sec:intro}
The Vera C. Rubin Observatory Legacy Survey of Space and Time (LSST) will make transient classification a problem of scale as well as depth. Its alert stream will be sparse, irregularly sampled, and survey-strategy dependent, with millions of alerts expected each night \citep{ivezic2019lsst}. Tidal disruption events (TDEs) arise when a star is disrupted by a massive black hole and provide a direct probe of black-hole demographics, accretion physics, and the nuclear environment \citep{1988Natur.333..523R}. \added{Forecasts suggest that LSST could detect approximately 35,000--80,000 TDEs over its ten-year survey, with the recovered yield depending on cadence and photometric selection criteria \citep{vanVelzen2011TDE,Bricman2020LSST,Bricman2023Strategy}.} Photometrically, however, TDEs overlap with several common contaminants, including AGN and other forms of nuclear variability, superluminous supernovae, and ordinary supernova subclasses, especially in the early alert history \citep{stein2024tdescore,bhardwaj2025photometric}.

% Time-domain astronomy is entering a regime in which alert volume, heterogeneity, and cadence complexity are becoming at least as important as raw depth. Rubin LSST is expected to generate millions of alerts per night, with a cadence that is sparse, irregular, band-dependent, and survey-strategy dependent \citep{ivezic2019lsst}. Within that context, tidal disruption events (TDEs) are an especially challenging target class. They are both intrinsically rare and astrophysically valuable, while also being frequently confused with a narrow set of photometric impostors that includes nuclear variability, superluminous supernovae, and several supernova subclasses \citep{1988Natur.333..523R,gezari2021tde,stein2024tdescore,bhardwaj2025photometric}.

Most existing photometric rare-transient pipelines rely on feature engineering followed by classical discriminative models. Here feature engineering refers to replacing the native time series with fitted or hand-defined summaries, such as rise and decline times, colors at a chosen phase, peak magnitudes, host offsets, or Gaussian-process hyperparameters. Recent examples include Gaussian-process-based feature extraction followed by boosted trees for simulated Rubin-like TDE classification \citep{bhardwaj2025photometric}, context-aware random-forest pipelines for rapid superluminous supernova (SLSN) discovery and TDE identification \citep{gomez2020fleet,gomez2023fleet}, image+metadata hybrid classifiers for real-time rare-transient identification \citep{sheng2024needle}, Zwicky Transient Facility (ZTF; \citealt{Bellm2019ZTF}) nuclear-transient TDE classifiers that combine engineered light-curve and host-context features with gradient-boosted trees \citep{stein2024tdescore}, and modern multimodal broker classifiers such as ATAT \citep{Cabrera2024} and ORACLE \citep{2025ApJ...995....4S} that learn from time-series and contextual inputs. Additional broker classifiers have been introduced in operational alert systems, including the ALeRCE light-curve classifier and its TDE expansion, as well as Fink classifier benchmarks for LSST-like alert streams \citep{SanchezSaez2020ALeRCE,PavezHerrera2025ALeRCETDE,Fraga2024Fink}. These methods differ in whether they fit explicit light-curve features, use image information, incorporate host-galaxy context, or learn directly from the alert history, but they share a practical tension. The most powerful predictors are often the hardest to define consistently across surveys, cadences, and partially observed light curves.

Neural processes (NPs) offer a different route. Conditional neural processes (CNPs) were introduced by \citet{garnelo2018conditional} as deterministic set-conditioned predictors, while the latent neural-process formulation of \citet{garnelo2018neural} added a global stochastic path for function-level uncertainty. Convolutional conditional neural processes \citep[ConvCNPs;][]{gordon2020convcnp} further introduced grid-based representations that let convolutional filters share information across nearby input locations. Together, these models define distributions over functions conditioned on context observations.
% They fit irregularly sampled, partially observed time series well and still permit amortized test-time inference. These features are attractive for astronomical light curves, where missingness is structured rather than incidental and where uncertainty under limited context is often scientifically central.
LSST alert light curves naturally fit this formulation because the available measurements form an irregular context set, and early classification can guide follow-up before the source evolution is fully observed.

Several recent astronomy studies have already explored this direction. \citet{kovacevic2023quasar} applied a conditional neural process framework to Rubin-oriented quasar light-curve modeling, while \citet{cvorovic2022agncnp} used CNPs for non-parametric modeling of AGN light curves. More recently, \citet{Chaini2026NightLANP} used attentive NPs for modeling supernova-like transient light curves, while \citet{Raju2026QNPY} developed a framework for AGN reverberation mapping. These works establish that NP-style models can handle stochastic, irregular astrophysical variability, but they do not directly address the classification problem posed by rare transient alerts in a multiband, multiclass, confuser-aware setting. NAPTIME is developed to address that classification gap.

In this work we present NAPTIME, the Neural Astrophysical Photometric Transient Identification and Modeling Engine. NAPTIME is a unified neural-process-based framework for photometric transient modeling and classification that supports both photometry-only and metadata-aware operation. This work develops a neural-process classifier for Rubin-like alert data, meaning sparse, irregular, multiband light curves with optional host context and incomplete early-time coverage. We use a family-level taxonomy, by which we mean a condensed set of broad astrophysical alert families formed by merging finer simulation classes into physically interpretable groups. The primary task is multiclass classification over this condensed Extended LSST Astronomical Time-series Classification Challenge 2 (ELAsTiCC2; \citealt{ELAsTiCC}) taxonomy, and the resulting posterior is interpreted both globally and through one-vs-rest TDE retrieval metrics under full-context and early-context conditions. Broad alert classification and TDE retrieval are treated as coupled tasks within a single model. The multiclass posterior defines the alert-family taxonomy, while its TDE component provides a rare-transient ranking score. The NAPTIME implementation and trained checkpoint releases are available online.\footnote{\url{https://github.com/nmearl/naptime}} \added{The version used in this work is archived on Zenodo under the MIT License at \dataset[doi:10.5281/zenodo.22082884]{https://doi.org/10.5281/zenodo.22082884} \citep{Earl2026NAPTIME}.}

Section~\ref{sec:background} reviews the photometric-classification and neural-process literature relevant to Rubin alert streams. Section~\ref{sec:data} describes the Many Artificial LSST Lightcurves based on Observations of Real Nuclear transients (MALLORN; \citealt{2026RASTI...5ag019M}) and ELAsTiCC2 simulated benchmarks together with the condensed ELAsTiCC2 classification taxonomy used in this work. Section~\ref{sec:method} presents the NAPTIME architecture, training objective, and inference setup. Section~\ref{sec:experiments} defines the experimental design, and Section~\ref{sec:results} reports the main benchmark, TDE-retrieval, and metadata-importance results. Section~\ref{sec:discussion} discusses limitations and interpretation, and Section~\ref{sec:conclusion} summarizes the main conclusions.

\section{Background and Related Work}\label{sec:background}
\subsection{Photometric classification of rare transients}
A major branch of transient-classification work has focused on deriving interpretable summary statistics from light curves and host context, then using tree-based models or related discriminative classifiers \citep[e.g.,][]{2019PASP..131k8002M,2021AJ....162..275B,2023ApJS..267...25H}. For TDEs, \citet{bhardwaj2025photometric} fit Gaussian processes (GPs) to ELAsTiCC2 light curves and extracted engineered features including rise and fade times, post-peak colors, color-evolution slopes, and GP hyperparameters. Their results showed that color and timescale information are particularly discriminative, with post-peak colors and GP length scales dominating feature importance. In real-survey ZTF data, \citet{stein2024tdescore} reached strong TDE precision and recall by combining hand-crafted light-curve features with extensive host and catalog context, including host colors, infrared variability, and external AGN catalog cross-matches.

Related work on other rare transients has reached similar conclusions. \citet{gomez2020fleet} designed a redshift-agnostic random-forest pipeline for SLSN selection that emphasized host separation, transient-host contrast, color near peak, and lightweight temporal summaries; \citet{gomez2023fleet} subsequently extended the same pipeline to TDE identification, finding that host-separation and transient color features are similarly discriminative. \citet{sheng2024needle} used a hybrid image+metadata architecture to identify rare transients from ZTF, showing that host information strongly boosts early classification, especially for source classes with strong environmental priors. This same host information also encodes the selection function of the training sample, and the classifier will tend to perform best for events in the most commonly represented host environments. Across these systems, high-utility classification signals are distributed across both the light curve and the source environment, and the value of contextual metadata generally grows as the photometric record becomes shorter or noisier.

These methods are strong baselines, but they also impose survey-specific design choices. Engineered features often require peak estimates, minimum numbers of measurements, or explicit host matching rules, and their behavior can change when cadence or signal-to-noise changes. Feature computation can also become expensive when it requires repeated light-curve fitting, host association, or catalog cross-matching across large alert streams \citep{SanchezSaez2020ALeRCE,Malanchev2021}. This emphasizes the need for a modeling framework that can ingest the native irregular measurements directly while still benefiting from contextual metadata when available.

\subsection{Neural processes and convolutional variants}
Conditional neural processes and latent neural processes were introduced in closely related work by \citet{garnelo2018conditional} and \citet{garnelo2018neural}, respectively, as models that combine the flexibility and uncertainty quantification of stochastic processes with the amortized inference of neural networks. Given a set of context observations, an NP predicts a distribution over target values at new input locations. In this setting, amortized inference means that a trained encoder--decoder network produces the predictive distribution directly for each new light curve, rather than refitting a separate model or solving a new optimization problem at test time. Unlike a Gaussian process, the model does not require explicit kernel specification or cubic-time inference at test time. Instead, an encoder maps context observations to a latent representation, and a decoder conditions on that representation to produce predictive distributions at target inputs.

CNPs remove the latent global variable and define a deterministic mapping from context sets to predictive distributions, which improves simplicity and efficiency but can limit global uncertainty modeling. ConvCNPs embed context observations on a grid and process that representation with convolutional layers. In the original ConvCNP formulation, translation equivariance means that shifting all input locations shifts the representation and predictions in the same way. For transient light curves, the useful property is more practical than literal: once irregular measurements are projected onto a regular relative-time grid, convolutional filters can reuse local temporal patterns at nearby phases. Qualitatively, this converts an irregular set of measurements into a representation on which nearby times can share information, so the model can learn reusable local motifs such as smooth rises, declines, and color evolution without requiring hand-defined peak times or cadence-specific features. For one-dimensional light curves, the practical value of this approach is not strict physical translation equivariance in absolute time, since transient evolution is anchored to event phase and survey window. Instead, the main advantage is that irregular observations are first projected onto a regular temporal grid, after which convolutional layers can learn local temporal signals such as rises, declines, and band-dependent shape changes.
%architectures; for one-dimensional time series, this provides an attractive inductive bias. Local patterns in time can be learned through convolution, while irregular observations are first projected into a representation defined over a common temporal support. 
Gaussian Neural Processes (GNPs) extend this family by making the stochastic predictive path more explicit and by tightening the connection between neural-process training and correlated predictive uncertainty \citep{Bruinsma2021GNP}.

This architecture is particularly relevant for astronomy because many survey light curves are irregularly sampled and partially observed, but still exhibit local temporal regularities such as smooth rises and post-peak declines, as well as band-dependent shape differences. ConvCNPs use a set representation whose output does not depend on the order in which observations are listed; the time coordinate remains part of each input point, so this does not mean that the light curve can be scrambled in time without consequence. The resulting representation retains the set-based flexibility of NPs while using the convolutional backbone to share information across nearby phases after the irregular observations have been projected to the grid. NAPTIME follows this broader neural-process lineage but is not a deterministic CNP or ConvCNP. It combines a convolutional temporal representation with a global latent path, placing it closer to a ConvGNP-style architecture adapted for irregular multiband transient classification.

\subsection{Neural processes in variable-source astronomy}
Several astronomy applications have used neural-process models for stochastic or irregular light curves, mostly outside the rare-transient classification setting. \citet{kovacevic2023quasar} used an upgraded conditional neural process framework for LSST-oriented quasar light curves, emphasizing that NPs can model uncertainty while avoiding the computational cost of Gaussian-process fitting on massive synoptic data sets. \citet{cvorovic2022agncnp} applied CNPs to AGN light curves as a non-parametric modeling alternative to damped random walk and GP approaches, arguing that NPs are well suited to sparse, gapped, and irregular light curves without explicit kernel assumptions. Related generative transient models such as ParSNIP also support the broader idea that reconstructive objectives can improve downstream classification and robustness \citep{2021AJ....162..275B}, while RAPID established the importance of explicit early-time classification under incomplete light-curve context \citep{2019PASP..131k8002M}. NightLANP provides a complementary neural-process application for class-agnostic Rubin light-curve reconstruction and calibration, with a focus distinct from the classification task studied here \citep{Chaini2026NightLANP}.

These studies motivate the use of NPs for light-curve modeling, but they stop short of the transient-classification setting that Rubin will make operationally urgent. Our work differs in two ways. First, the model is trained to perform multiclass classification across a broader family-level alert taxonomy while still reconstructing or interpolating the light curve. Second, the evaluation is explicitly designed around partial context, including sequential prefix sweeps intended to mimic progressively increasing alert information. In a prefix sweep, only the earliest fraction of each object's retained photometric measurements is used, so the evaluation directly probes how performance scales with the accumulated alert history. NAPTIME uses the NP as both a regression model and a probabilistic backbone for alert-facing classification.

\section{Data}\label{sec:data}
\subsection{MALLORN as a photometry-only benchmark}
We use the MALLORN challenge data as a photometry-only benchmark for TDE classification. \added{The labeled light curves and object table were obtained from the official MALLORN Astronomical Classification Challenge data release.\footnote{\url{https://www.kaggle.com/competitions/mallorn-astronomical-classification-challenge/data}}} MALLORN provides irregular multiband light curves with redshift information but lacks the host-galaxy context available in ELAsTiCC2. This construction tests the photometric backbone under a different simulation design, not only under a reduced metadata setting. MALLORN light curves are derived from real ZTF nuclear-transient observations forward-modeled into LSST-like bands, cadence, and noise properties, with TDE examples drawn from spectroscopically classified ZTF events. This construction differs from the PLAsTiCC/ELAsTiCC simulation lineage and includes information from the ZTF TDE sample that was not available to the earlier challenges. For the results reported here, we use the labeled MALLORN taxonomy directly and interpret the TDE probability one-vs-rest from the multiclass posterior rather than training a separate binary classifier. MALLORN serves as a secondary benchmark for photometry-only performance and robustness to a different procedure for assembling transients and contaminants.

\subsection{ELAsTiCC2 as the Rubin-like benchmark}
Our primary experiments use the ELAsTiCC2 simulated Rubin-like training sample. \added{We use the \texttt{ELASTICC2\_TRAIN\_02} SNANA FITS release distributed through the DESC ELAsTiCC2 data portal.\footnote{\url{https://portal.nersc.gov/cfs/lsst/DESC_TD_PUBLIC/ELASTICC/ELASTICC2_TRAINING_SAMPLE_2}}} ELAsTiCC2 provides
%SNANA-style \texttt{HEAD}/\texttt{PHOT} files with 
synthetic observations with realistic cadence, noise, host associations, and updated transient models intended to better reflect Rubin-era expectations. ELAsTiCC2 TDE light curves are generated from SNANA simulation templates \citep{Kessler2009SNANA} calibrated to observed TDE properties. The class frequencies and simulated source populations are chosen to make the benchmark useful for training and evaluation; they should not be read as predictions for the Rubin alert-stream class mixture. For TDEs in particular, existing light-curve templates and host-galaxy models do not yet cover the full observed diversity, including early bumps, broad color diversity, and the wide range of late-time decline, plateau, and rebrightening behavior seen in real events \citep{gezari2021tde}. Because the present work is aimed at a Rubin-oriented classification framework, ELAsTiCC2 serves as the central benchmark for both the multiclass classification experiments and the sequential prefix sweeps. ELAsTiCC2 builds on the earlier PLAsTiCC classification challenge by adding a more broker-oriented simulation structure with updated source models, host associations, and alert-stream realism \citep{2023ApJS..267...25H,ELAsTiCC}.

Each object is represented by irregularly sampled photometric measurements in Rubin bands, together with challenge-supplied contextual metadata such as Milky Way reddening, redshift-related quantities, host separation metrics, including the directional light radius (dDLR:  a host-normalized offset measure commonly used in transient-host association), host stellar-mass proxies, and host magnitudes. It is important to note that the class counts reflect simulation design, not intrinsic volumetric rates. ELAsTiCC2 allocates a large number of TDE examples to ensure adequate coverage. Some schema-defined metadata columns are unpopulated in the training files used here; these are removed from the active metadata vector during preprocessing.

\subsection{Condensed ELAsTiCC2 classification taxonomy}
Rather than training on the full ELAsTiCC2 model inventory, we collapse the release directories into the 15 physically motivated alert families listed in Table~\ref{tab:elasticc-taxonomy}. We refer to this as our classification taxonomy throughout the rest of the paper.

\begin{deluxetable*}{cll}
\tablecaption{Condensed ELAsTiCC2 classification taxonomy used in this work.\label{tab:elasticc-taxonomy}}
\tablehead{\colhead{ID} & \colhead{Alert family} & \colhead{Merged ELAsTiCC2 source classes}}
\startdata
1 & TDE & TDE \\
2 & AGN-like nuclear variability & CLAGN \\
3 & Ca-rich/fast transients & CART \\
4 & Intermediate-luminosity optical transients (ILOTs) & ILOT \\
5 & Kilonovae & KN\_B19, KN\_K17 \\
6 & Superluminous supernovae (SLSNe) & \begin{tabular}[t]{@{}l@{}} SLSN-I+host, SLSN-I\_no\_host \end{tabular} \\
7 & Pair-instability supernovae (PISNe) & PISN \\
8 & SN~Ia & \begin{tabular}[t]{@{}l@{}} SNIa-SALT3, SNIa-91bg, SNIax \end{tabular} \\
9 & SN~II & \begin{tabular}[t]{@{}l@{}} SNII+HostXT\_V19, SNII-NMF,\\ SNII-Templates, SNIIn+HostXT\_V19,\\ SNIIn-MOSFIT \end{tabular} \\
10 & Stripped-envelope supernovae (SN~Ibc) & \begin{tabular}[t]{@{}l@{}} SNIIb+HostXT\_V19, SNIb+HostXT\_V19,\\ SNIb-Templates, SNIc+HostXT\_V19,\\ SNIc-Templates, SNIcBL+HostXT\_V19 \end{tabular} \\
11 & M-dwarf flares & Mdwarf-flare \\
12 & Pulsators & Cepheid, RRL, d-Sct \\
13 & Eclipsing binaries & EB \\
14 & Dwarf novae & dwarf-nova \\
15 & Microlensing events & \begin{tabular}[t]{@{}l@{}} uLens-Binary,\\ uLens-Single-GenLens,\\ uLens-Single\_PyLIMA \end{tabular} \\
\enddata
\end{deluxetable*}

This taxonomy is broad enough to function as a realistic broker-style alert benchmark, but still coarse enough to remain physically interpretable. In ELAsTiCC2, the nuclear-variability bucket is generated using changing-look AGN \citep[CLAGN;][]{matt2003,lamassa2015}, so we refer to that family conservatively as \emph{AGN-like} instead of generic AGN. The resulting task lets us evaluate NAPTIME both as a family-level classifier and, through one-vs-rest metrics, as a TDE-oriented ranking model embedded in a broader alert stream.

\section{Methodology}\label{sec:method}
\subsection{Problem setup}
For each transient, we observe a set of irregularly sampled multiband measurements
\begin{equation}
\mathcal{D} = \{(t_i, b_i, f_i, \sigma_i)\}_{i=1}^{N},
\end{equation}
where $t_i$ is time, $b_i$ is the filter index, $f_i$ is flux, and $\sigma_i$ is the flux uncertainty. Optional metadata are represented as a vector $m$ together with an elementwise availability mask. The task is to infer both a predictive distribution for held-out photometric targets and a class posterior over a collapsed taxonomy of transient classes.

The neural-process framing splits each light curve into a context set $\mathcal{C}$ and a target set $\mathcal{T}$. During training, $\mathcal{C}$ contains only a fraction of the observed points, while $\mathcal{T}$ contains the held-out points used for reconstruction. The model must therefore reconstruct the withheld photometry and classify the source from partial context within a single training objective. This setup encourages the learned representation to remain useful even when the number and temporal placement of observations vary. At evaluation time we consider both random partial-context splits and sequential prefix splits in which only the earliest observations are retained.

\begin{figure}
\centering
\includegraphics[width=\columnwidth]{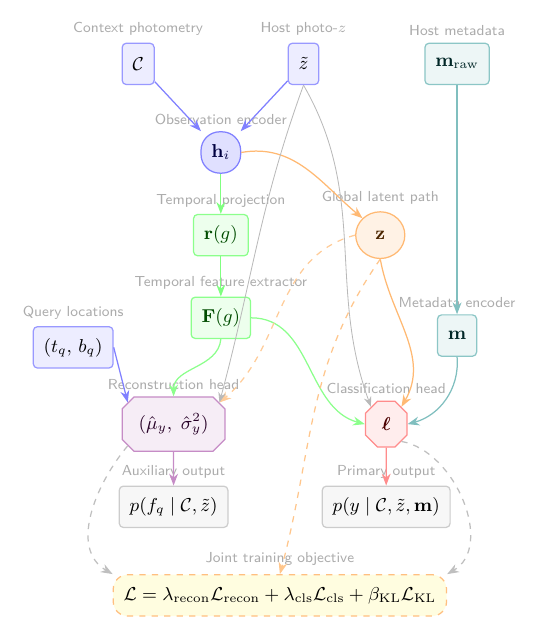}
\caption{Symbolic architecture of NAPTIME. Context photometry $\mathcal{C}$ and host photo-$z$ are encoded into per-observation features $\mathbf{h}_i$, projected to a regular temporal representation $\mathbf{r}(g)$, and processed into backbone features $\mathbf{F}(g)$. Filter identities enter the point encoder and decoder through learned band embeddings. A global latent path produces $\mathbf{z}$, host metadata are encoded as $\mathbf{m}$, and the model jointly produces a predictive flux distribution and a family-level class posterior under the training objective shown at bottom. The variables and operations shown here are defined in Section~\ref{sec:method}.}
\label{fig:naptime-arch}
\end{figure}

\subsection{Input normalization and representation}
%Within each object, times are shifted and rescaled to the unit interval based on the observed span of the light curve. 
Times are normalized to the observed span of each object, so the temporal backbone learns relative light-curve structure as opposed to absolute physical timescales. \added{For each object, the temporal interval is defined by the earliest and latest retained photometric epochs after selecting the six supported bands and requiring finite fluxes and finite, positive uncertainties. No signal-to-noise threshold is applied, and low-significance and negative-flux measurements are retained.} Absolute duration information is reintroduced only through summary features such as $\log \Delta t_{\mathcal{C}}$ at the classifier stage. Fluxes and flux uncertainties are normalized bandwise using robust statistics computed from the training set. Because challenge data contain outliers and occasionally extreme scale differences, we enforce a minimum scale floor and clip normalized fluxes and uncertainties to finite ranges. Normalized fluxes are clipped to $[-200,\,200]$ times the per-band scale, and uncertainties to $[10^{-3},\,200]$; these limits leave the dynamic range of well-measured sources unaffected and in practice suppress only extreme outliers. This improves numerical stability without the need to mask valid observations. For the broker-style ELAsTiCC2 experiments reported here, the dedicated scalar redshift input is the host-galaxy photometric redshift, and related distance or redshift estimates are excluded from the auxiliary metadata vector to avoid duplicating the same information. Redshift is normalized separately to a compact range, while metadata are standardized fieldwise and masked where unavailable.

The model ingests each context point as a tuple consisting of normalized time, normalized flux, normalized flux uncertainty, band identity, and optionally normalized redshift. Band identity is represented with a learned embedding, and time is expanded using a Fourier feature map so that the downstream encoder can capture both smooth trends and sharper temporal structure. Global summary quantities such as the observed temporal span and the number of context points are also propagated to the classifier head.

Concretely, each observed point is mapped to the encoded feature shown in Figure~\ref{fig:naptime-arch},
\begin{equation}
\mathbf{h}_i = m_i \cdot \mathrm{MLP}\!\left([f_i,\log \sigma_i,\mathbf{e}_{b_i},\bm{\tau}(t_i),\tilde{z}]\right),
\end{equation}
where $\mathbf{e}_{b_i}$ is the learned band embedding, $\bm{\tau}(t_i)$ is a Fourier time embedding, $\tilde{z}$ is the normalized host photo-$z$, and $m_i$ is the observation mask used to zero padded entries. The resulting sequence $\{\mathbf{h}_i\}$ is the basic pointwise representation used by both the temporal path and the latent path. Flux uncertainties enter as $\log \sigma_i$ because photometric noise spans several orders of magnitude across the survey. The log transform gives the encoder a uniform and numerically well-conditioned input scale.

\subsection{ConvGNP-style neural-process backbone}
NAPTIME is built around a one-dimensional ConvGNP-style architecture \citep{gordon2020convcnp,Bruinsma2021GNP}. The encoded point features $\mathbf{h}_i$ defined above are projected onto a common temporal grid using a Gaussian set convolution. This step behaves as a continuous-to-grid interpolation stage, allowing the model to combine irregularly sampled inputs into a fixed-resolution sequence while remaining invariant to the order in which context points are listed.

For grid location $g$ and set-convolution scale $s$, we write
\begin{equation}
w_{ig}^{(s)} = \exp\!\left[-\frac{(t_i-g)^2}{2s^2}\right] m_i,
\end{equation}
\begin{equation}
\mathbf{r}^{(s)}(g) = \frac{\sum_i w_{ig}^{(s)} \mathbf{h}_i}{\sum_i w_{ig}^{(s)}},
\end{equation}
with an accompanying density channel $\rho^{(s)}(g)=\sum_i w_{ig}^{(s)}$. Concatenating the multiscale representations $\mathbf{r}^{(s)}(g)$ and density channels across scales yields the temporal projection block denoted $\mathbf{r}(g)$ in Figure~\ref{fig:naptime-arch}. A $1\times 1$ convolution then maps this multiscale stack to the grid feature channels used by the temporal backbone.

The gridded representation is processed by a convolutional backbone that learns local temporal patterns such as smooth rises, extended fades, and band-dependent shape differences. The resulting backbone features are denoted $\mathbf{F}(g)$ in Figure~\ref{fig:naptime-arch}. A decoder then predicts, for each target input location and band, a Gaussian mean and variance for the normalized flux. Measurement uncertainties are reincorporated in the likelihood by adding the reported photometric variance to the model variance.

Given a query time-band pair $(t_q,b_q)$, we linearly interpolate the backbone features to obtain $\tilde{\mathbf{F}}_q$ and decode
\begin{equation}
[\hat{\mu}_y,\tilde{v}] = \mathrm{MLP}\!\left([\tilde{\mathbf{F}}_q,\mathbf{e}_{b_q},\bm{\tau}(t_q),\tilde{z},\mathbf{z}]\right),
\end{equation}
\begin{equation}
\hat{\sigma}_y^2 = \left(\mathrm{softplus}(\tilde{v}) + \epsilon\right)^2.
\end{equation}
Here $\mathbf{z}$ is the global latent vector defined in the following subsection, which captures object-level variability not expressed by the local convolutional backbone, and at inference time it is replaced by the posterior mean $\bm{\mu}_z$. This defines the predictive distribution shown in Figure~\ref{fig:naptime-arch},
\begin{equation}
p(f_q \mid \mathcal{C}, t_q, b_q, \tilde{z}, \mathbf{z})
= \mathcal{N}\!\left(\hat{\mu}_y,\hat{\sigma}_y^2 + \sigma_q^2\right).
\end{equation}
Here $\sigma_q$ is the reported photometric uncertainty of the target measurement, so observational noise is included explicitly in the reconstruction likelihood.

\subsection{Global latent path}
To retain the probabilistic character of the neural-process family, we include a global latent path inspired by the latent-variable NP formulation of \citet{garnelo2018neural}. A separate encoder summarizes the context set into a latent mean and log-variance, defining an object-level latent variable that captures global aspects of the light curve not easily represented by local convolution alone. During training, the latent variable is sampled in differentiable form as $\mathbf{z}=\bm{\mu}_z + \bm{\sigma}_z \odot \bm{\epsilon}$ with $\bm{\epsilon} \sim \mathcal{N}(\mathbf{0}, I)$; during evaluation, the predictive path uses the latent mean $\bm{\mu}_z$. A Kullback--Leibler penalty regularizes the latent posterior toward a prior, with optional warm-up to stabilize optimization.

The latent path operates directly on the pointwise encoded context features as opposed to the temporal grid. Let
\begin{equation}
\bar{\mathbf{h}} = \frac{\sum_i m_i \mathbf{h}_i}{\sum_i m_i},
\qquad
\hat{\mathbf{h}} = \max_{\{i:m_i=1\}} \mathbf{h}_i,
\end{equation}
and define $\mathbf{r}_{\mathrm{lat}}=[\bar{\mathbf{h}},\hat{\mathbf{h}}]$. The latent encoder produces
\begin{equation}
(\bm{\mu}_z,\log \bm{\sigma}_z^2) = \mathrm{MLP}(\mathbf{r}_{\mathrm{lat}}),
\end{equation}
which defines the approximate posterior
\begin{equation}
q(\mathbf{z}\mid\mathcal{C}) = \mathcal{N}\!\left(\bm{\mu}_z,\mathrm{diag}(\bm{\sigma}_z^2)\right).
\end{equation}
During training, we sample $\mathbf{z}$ with the differentiable parameterization above; at inference time, we use the posterior mean $\bm{\mu}_z$.

%This latent path turned out to be important in practice. Earlier deterministic variants of the model underperformed stronger latent versions, and reducing the reconstruction objective substantially degraded performance. 
The latent path is included to preserve the probabilistic character of the model family and to provide an object-level summary that is not easily captured by local convolution alone. Because this path is stochastic during training, it can also represent object-level diversity not captured by the deterministic convolutional grid. In exploratory development, deterministic variants were generally less stable and less effective under sparse context, so we retain the latent path in the final architecture. Empirically, the latent generative path improves ranking quality and stabilizes early-context behavior. The resulting model is best described as a probabilistic neural process with a classifier head.

\subsection{Metadata branch}
When metadata are enabled, the model uses a dedicated metadata encoder that ingests standardized metadata values together with an availability mask. The mask is essential because many challenge-provided metadata fields are missing for subsets of objects, and the same issue will occur in real alert streams when host association, redshift estimation, or external catalog coverage is incomplete. Metadata are not used to replace the photometric backbone; instead, they provide a parallel contextual stream that is fused only near the classifier head. This design allows the same model to operate in both photometry-only and metadata-aware regimes, which is important for comparing MALLORN and ELAsTiCC2 without changing the fundamental architecture.

If $\mathbf{m}_{\mathrm{raw}}$ denotes the standardized metadata vector and $\mathbf{v}$ its availability mask, then the metadata encoder produces the compact representation shown in Figure~\ref{fig:naptime-arch},
\begin{equation}
\mathbf{m} = \mathrm{MLP}\!\left([\mathbf{m}_{\mathrm{raw}},\mathbf{v}]\right).
\end{equation}

The metadata branch is particularly relevant for source classes with strong environmental priors. In the family-level taxonomy, this is most obvious for TDEs and AGN-like nuclear variability, where host separation and related nuclearity indicators can be informative, but the same branch also helps separate the supernova and fast-transient families through host magnitude and mass information. A classifier trained on these simulations may consequently be less sensitive to TDEs in atypical host environments or at large nuclear offsets.

\subsection{Classification head}
For classification, the convolutional grid representation is summarized through global pooling operations, and the resulting pooled features are concatenated with optional metadata embeddings, redshift summaries, context-size summaries, and the latent posterior parameters. This combined vector is processed by a multilayer perceptron to produce class logits. The classifier can therefore access both local morphological information encoded in the grid and object-level contextual information from the latent and metadata paths.

In practice, we combine mean-pooled, max-pooled, and latent-attended summaries of $\mathbf{F}(g)$, then append $\tilde{z}$, the context span $\log \Delta t_{\mathcal{C}}$, the context size $\log N_{\mathcal{C}}$, the latent posterior parameters $(\bm{\mu}_z,\log\bm{\sigma}_z^2)$, and the metadata embedding $\mathbf{m}$ when present. Writing the fused feature vector as $\mathbf{p}$, the classifier head shown in Figure~\ref{fig:naptime-arch} produces
\begin{equation}
\bm{\ell} = \mathrm{MLP}(\mathbf{p}),
\qquad
p(y\mid\mathcal{C},\tilde{z},\mathbf{m}) = \mathrm{softmax}(\bm{\ell}).
\end{equation}

We use a multiclass cross-entropy objective for the ELAsTiCC2 family taxonomy, with inverse-frequency class weights derived from the training distribution. These prevent the loss from being dominated by the large SN~Ia and SN~II populations during training, and the weighting can be adjusted for a different operating class distribution.
%with class weights derived from the training distribution. 
For the MALLORN results reported here, the classifier uses the MALLORN multiclass taxonomy and the TDE score is interpreted one-vs-rest from that multiclass posterior, analogous to the ELAsTiCC2 analysis. The implementation remains a single NP framework whose task definition changes with the target taxonomy.

\subsection{Joint training objective}
The full loss is
\begin{equation}
\mathcal{L} = \lambda_{\rm recon}\,\mathcal{L}_{\rm recon} + \lambda_{\rm cls}\,\mathcal{L}_{\rm cls} + \beta_{\rm KL}\,\mathcal{L}_{\rm KL},
\end{equation}
where $\mathcal{L}_{\rm cls}$ is the classification loss and $\mathcal{L}_{\rm KL}$ regularizes the latent path. \added{The coefficients $\lambda_{\rm recon}$ and $\lambda_{\rm cls}$ set the relative contributions of reconstruction and classification, while $\beta_{\rm KL}$ controls the strength of latent regularization. We use $\lambda_{\rm recon}=\lambda_{\rm cls}=1$ and increase $\beta_{\rm KL}$ from zero to $5\times10^{-4}$ over the first 20 epochs.} For held-out target photometry, the reconstruction term is the Gaussian negative log-likelihood
\begin{equation}
\mathcal{L}_{\rm recon}
= \frac{1}{2}\sum_q \left[
\frac{(f_q-\hat{\mu}_{y,q})^2}{\hat{\sigma}_{y,q}^2+\sigma_q^2}
+
\log\!\left(\hat{\sigma}_{y,q}^2+\sigma_q^2\right)
\right],
\end{equation}
up to an additive constant. 
%Reducing or removing the reconstruction term materially worsened classification. The generative objective therefore contributes useful inductive bias under sparse context by encouraging the model to preserve physically relevant light-curve structure instead of focusing only on decision boundaries.
The reconstruction term encourages the shared representation to preserve physically informative light-curve structure while learning the classification task, which is especially useful when only sparse context is available.

\section{Experimental Design}\label{sec:experiments}
We perform three sets of tests. First, we evaluate performance in a photometry-only regime. Second, we measure the added value of contextual metadata on Rubin-like challenge data. Third, we measure how classification quality changes as observational context accumulates.

MALLORN serves as a secondary benchmark for the photometry-dominant regime, while ELAsTiCC2 remains the primary benchmark for metadata-aware and broker-style family classification. The main ELAsTiCC2 experiment uses a metadata-aware 15-family model and a matched photometry-only comparison with the same architecture and optimization settings. We report macro F1, weighted F1, macro area under the receiver operating characteristic curve (AUROC), and top-1 accuracy, together with per-class metrics and confusion matrices. Precision is defined as $P = \mathrm{TP}/(\mathrm{TP}+\mathrm{FP})$ and recall as $R = \mathrm{TP}/(\mathrm{TP}+\mathrm{FN})$, where $\mathrm{TP}$, $\mathrm{FP}$, and $\mathrm{FN}$ denote true positives, false positives, and false negatives. The F1 score is the harmonic mean of precision and recall,
\begin{equation}
\mathrm{F1} = \frac{2PR}{P+R},
\end{equation}
so it penalizes classifiers that achieve high purity at the expense of missed events, or high completeness at the expense of many false alarms. Macro F1 gives equal weight to each class and is sensitive to performance on rare families such as TDEs and kilonovae, while weighted F1 averages the same classwise scores with weights set by class frequency and emphasizes the bulk alert population. AUROC measures how well the model ranks positives above negatives as the decision threshold is varied, independent of any single operating point. Top-1 accuracy reports the fraction of objects whose most probable predicted class matches the true label and is included because it is easy to interpret operationally, even though it is less informative than macro F1 for strongly imbalanced taxonomies. For TDE-focused interpretation we additionally evaluate one-vs-rest TDE retrieval through average precision and threshold-dependent precision--recall behavior. For both ELAsTiCC2 models, this TDE score is read from the same multiclass posterior, so the photometry-only ELAsTiCC2 model supplies both a broad family classifier and a TDE ranking model without retraining a separate binary classifier. For early classification we perform sequential prefix sweeps in which only the earliest fraction of retained photometric measurements in each light curve is used. This is more informative for alert triage with progressively accumulating Rubin observations than a random partial-context split, although we also use random partial contexts during training and for auxiliary evaluation.

\section{Results}\label{sec:results}
The central result is that NAPTIME performs strongly on the Rubin-like ELAsTiCC2 family benchmark and that metadata yields a measurable gain. On the matched full 15-family ELAsTiCC2 experiment, using all available training objects rather than a size-capped subset, the metadata-aware model reaches a macro F1 of 0.9026 and a macro AUROC of 0.9906. The matched photometry-only model reaches 0.8741 and 0.9862. Thus metadata improves macro F1 by 0.0284 and macro AUROC by 0.0044, while also improving weighted F1 and top-1 accuracy. Table~\ref{tab:elasticc2-large} summarizes the matched comparison.

\begin{deluxetable*}{llcccc}
\tablecaption{Primary benchmark results for NAPTIME on the full ELAsTiCC2 family task and the secondary MALLORN photometry-only benchmark. MALLORN is not a family-level ELAsTiCC2 analogue; it is included as a reference photometry-only benchmark without a taxonomy-matched comparison to ELAsTiCC2.\label{tab:elasticc2-large}}
\tablehead{\colhead{Benchmark} & \colhead{Model} & \colhead{Macro F1} & \colhead{Weighted F1} & \colhead{Macro AUROC} & \colhead{Top-1 Accuracy}}
\startdata
ELAsTiCC2 & Families + metadata & 0.9026 & 0.8856 & 0.9906 & 0.8859 \\
ELAsTiCC2 & Families photometry only & 0.8741 & 0.8624 & 0.9862 & 0.8632 \\
MALLORN & Photometry only & 0.6926 & \nodata & 0.9579 & \nodata \\
\enddata
\end{deluxetable*}

\begin{figure*}
\centering
\includegraphics[width=\textwidth]{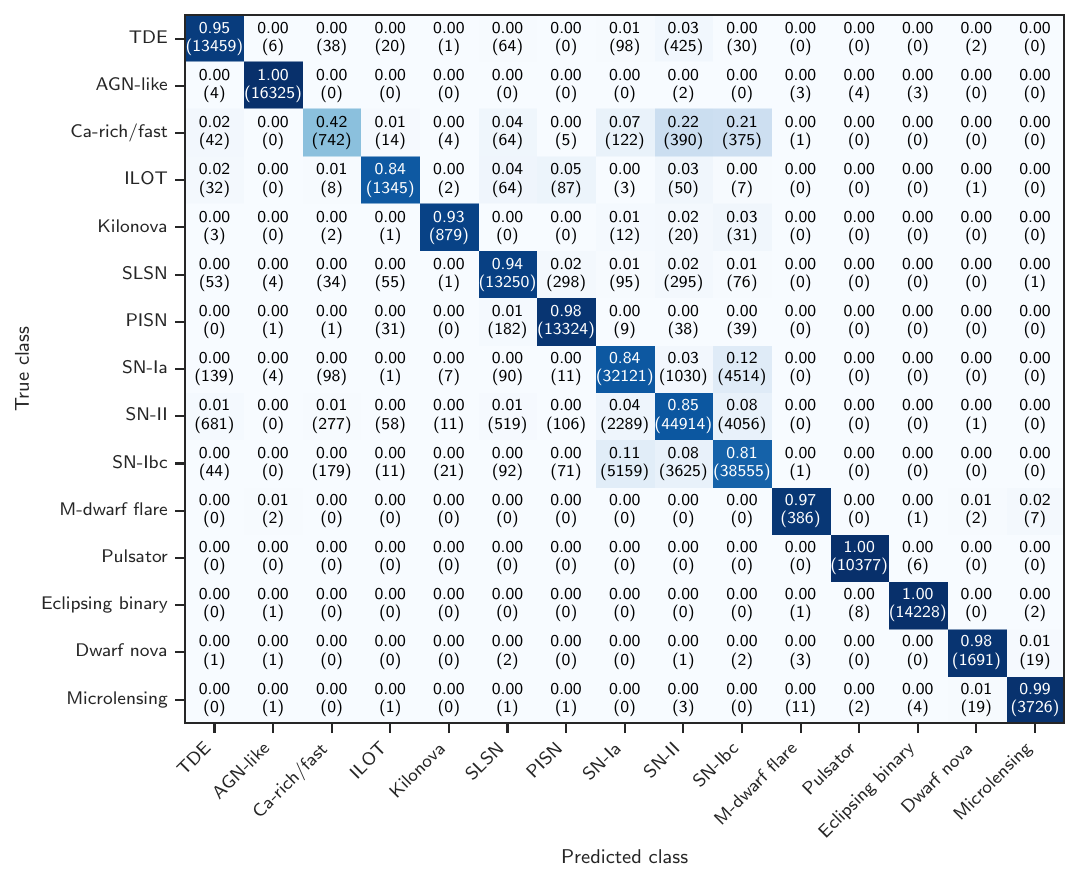}
\caption{Row-normalized confusion matrix for the metadata-aware family-level ELAsTiCC2 model. The dominant remaining confusions occur within the transient families as opposed to between TDEs and the bulk of the alert taxonomy.}
\label{fig:cm}
\end{figure*}

The improvement is not concentrated in an already easy class. In matched evaluations, metadata most strongly improves the harder transient families, including Ca-rich/fast events, ILOTs, PISNe, kilonovae, and the core supernova families, while the already well-separated periodic classes remain close to saturation. The periodic and recurrent families (pulsators, eclipsing binaries, M-dwarf flares, and dwarf novae) reach $\mathrm{F1} > 0.96$ in both models. Their period or rapidly recurrent light-curve structure is strongly separable from the smooth rises and declines of explosive transients, and the distinction persists even when only a fraction of the light curve has been observed.
 % This result suggests that the metadata does not only encode the dominant class label, but helps resolve the astrophysical boundaries that matter most for alert triage. 
 This result suggests that the metadata contributes a non-trivial boost to non-dominant classes and helps resolve some of the astrophysical boundaries that matter most for alert triage.
 Figure~\ref{fig:cm} shows the full multiclass confusion structure of the metadata-aware model. The remaining confusions identify the family boundaries that are still most difficult.

\added{The Ca-rich/fast family has the lowest recall among the transient families, at 0.422. Ca-rich events are identified most securely through their rapid transition to the nebular phase and strong calcium emission, neither of which is represented explicitly in broadband light curves. Near maximum light, they can resemble Type~Ib/c SNe, and some observed objects bridge the Ca-rich and stripped-envelope/core-collapse populations \citep{2012ApJ...755..161K,2017ApJ...846...50M}. The ELAsTiCC2 CaRT models likewise overlap the photometric parameter space of radioactive SN~Ibc explosions, while the grouped SN~II family contains several heterogeneous template populations. The smaller Ca-rich/fast training sample compounds this physical and simulated overlap. Reciprocal confusion from SN~II and SN~Ibc into the Ca-rich/fast class is present but visually suppressed by the larger sizes of those classes in the row-normalized matrix.}

\added{The simulated host-offset distributions provide little additional separation. Although observed Ca-rich transients frequently occur at large projected offsets \citep{2012ApJ...755..161K}, the ELAsTiCC2 CaRT population does not show an enhanced large-offset tail relative to the grouped SN~II and SN~Ibc populations. The metadata provide limited leverage for resolving these families.}

\begin{figure}
\centering
\includegraphics[width=0.95\columnwidth]{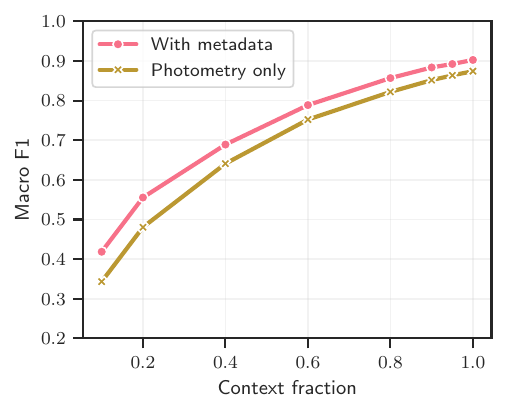}
\caption{Sequential prefix-sweep performance on the family-level ELAsTiCC2 benchmark. The metadata-aware model consistently outperforms the photometry-only comparison model, with the largest gains at the earliest context fractions.}
\label{fig:prefix}
\end{figure}

The sequential prefix sweeps show the same pattern. Using only the earliest 10\% of retained photometric measurements in each light curve, the metadata-aware family-level ELAsTiCC2 model reaches macro F1 $\approx 0.42$, while the photometry-only model reaches $\approx 0.34$. For the held-out ELAsTiCC2 objects, this 10\% prefix corresponds to a median of 18 observations spanning 36 days, with an interquartile span of 23--56 days. The metadata gain persists across all context fractions and is largest in the low-context regime, which is consistent with host information being most useful before the light curve itself becomes more informative. Figure~\ref{fig:prefix} shows that the metadata-aware model maintains an advantage across the full prefix range, with the largest separation at the smallest context fractions.

\begin{figure}
\centering
\includegraphics[width=\columnwidth]{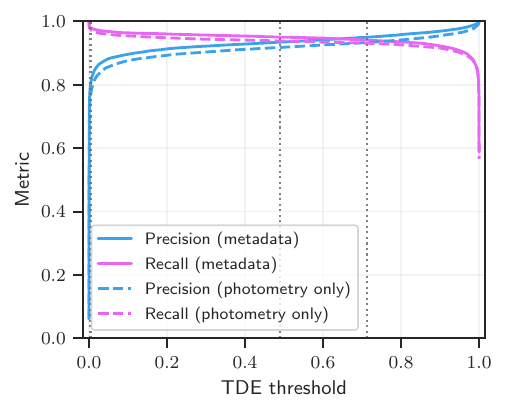}
\caption{TDE one-vs-rest precision and recall as functions of TDE score threshold on the family-level ELAsTiCC2 benchmark. Vertical markers indicate the three operating thresholds used in Figure~\ref{fig:tde-binary-cm}: approximately 80\% precision, 95\% precision, and 95\% recall.}
\label{fig:tde-threshold}
\end{figure}

\begin{figure*}
\centering
\includegraphics[width=0.98\textwidth]{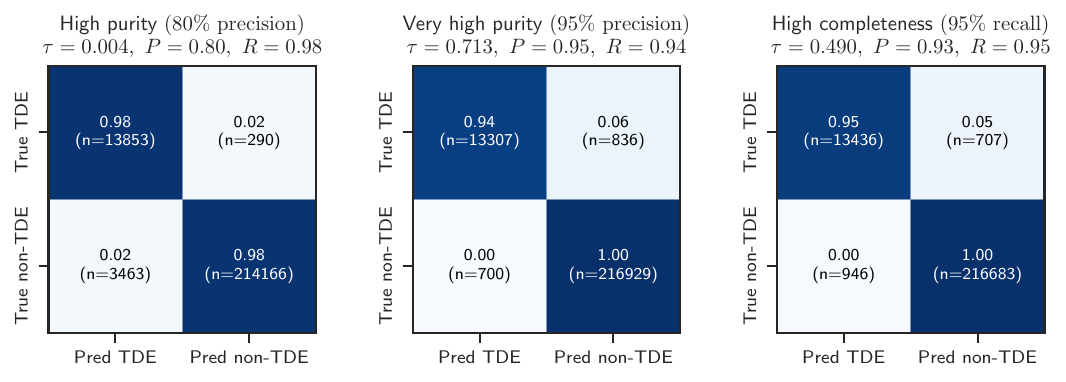}
\caption{Thresholded TDE-versus-non-TDE confusion matrices for the metadata-aware family-level ELAsTiCC2 model at three comparison operating points: approximately 80\% precision, 95\% precision, and 95\% recall. Each panel is row-normalized and annotated with raw counts. Displayed fractions are rounded to two decimal places.}
\label{fig:tde-binary-cm}
\end{figure*}

For TDE-focused interpretation, the family-level posterior can be viewed one-vs-rest. In the matched family-level ELAsTiCC2 evaluation, the metadata-aware model reaches TDE average precision 0.985, compared with 0.979 for the photometry-only variant. At threshold 0.5, precision is 0.935 with metadata and 0.919 without metadata, while recall rises from 0.938 to 0.950 when metadata are included. The high-purity operating points are more revealing. At precision $\approx 0.90$, recall increases from 0.945 without metadata to 0.963 with metadata, and at precision $\approx 0.95$ it increases from 0.919 to 0.941. Figure~\ref{fig:tde-threshold} shows the threshold-dependent precision-recall trade-off for TDE retrieval. To highlight the operating-point consequences, Figure~\ref{fig:tde-binary-cm} shows thresholded binary TDE-versus-non-TDE confusion matrices at operating points of approximately 80\% precision, 95\% precision, and 95\% recall, motivated by \citet{bhardwaj2025photometric}.

\subsection{Permutation Importance of Metadata}
\begin{figure}
\centering
\includegraphics[width=0.96\columnwidth]{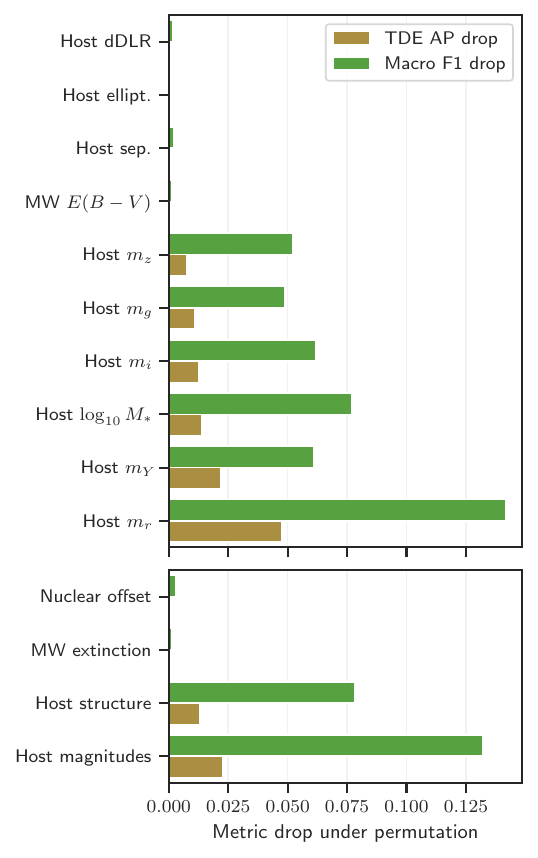}
\caption{Permutation-importance analysis for the metadata-aware ELAsTiCC2 family classifier. The top panel shows individual metadata fields and the lower panel shows grouped metadata families. Bars report the drop in TDE average precision and macro F1 under permutation. Larger bars indicate stronger model reliance on that metadata field or group.}
\label{fig:permimp}
\end{figure}

To clarify which contextual inputs the broker-style family classifier actually uses, we performed a permutation-importance analysis on the held-out ELAsTiCC2 validation set. Figure~\ref{fig:permimp} summarizes the performance drop induced by permuting either individual metadata fields or grouped metadata families while leaving the trained model fixed. In this benchmark, the largest independent contributions come from host photometric and structural quantities.
% Perhaps surprisingly, the analysis finds that the independent contribution of the nuclear-offset measurements is small after conditioning on the other metadata,
%little significance in the nuclear-offset measurements, instead finding that the dominant signals are host photometric and structural quantities. 
At the group level, host magnitudes produce the largest degradation, with a macro-F1 drop of $\sim$0.13 and a TDE-average-precision drop of $\sim$0.02, followed by host structure with drops of $\sim$0.08 and $\sim$0.01. At the field level, host $r$-band magnitude is the strongest single feature, followed by host stellar mass and the redder host magnitudes. Host magnitudes are proxies for stellar mass and, through apparent brightness, for distance. In the ELAsTiCC2 validation set, the host-magnitude fields have correlation ratios with the class label of $\eta=0.46$--0.53 and Spearman correlations with host photo-$z$ of $\rho=0.51$--0.58. Their high permutation importance therefore cannot be attributed to a single astrophysical cause; it reflects a mixture of host-environment information and distance-dependent sample structure.
%In contrast, the independent contribution of host separation and dDLR is weak in this full family-level benchmark. 
In contrast, host separation and dDLR make only a small additional contribution once the broader metadata vector is already available. 
%This suggests that when the broader host context is available, the trained classifier relies more strongly on host environment and photometric host properties than on nuclear-offset metadata alone.
This may indicate redundancy with other host features, but it may also reflect limitations of the simulated offset distributions in ELAsTiCC2, so the result should not be taken as a general statement that nuclearity is uninformative for real TDE classification. \added{Optically selected TDEs are predominantly nuclear \citep{gezari2021tde,Yao2023TDEDemographics}, although the offset events AT2024tvd and TDE~2025abcr demonstrate that a small off-nuclear population exists \citep{Yao2025OffsetTDE,Stein2026OffsetTDE}. Core-collapse SNe span a wider range of host offsets, so accurate astrometric separations may prove more discriminative in observed alert streams than these simulations suggest.}

MALLORN serves as a useful photometry-dominant complement. On the full labeled class inventory available in that dataset, NAPTIME reaches macro $\mathrm{F1} = 0.6926$ and macro $\mathrm{AUROC} = 0.9579$, while TDE one-vs-rest average precision reaches 0.730. The lower MALLORN score likely reflects several differences from ELAsTiCC2: the absence of host metadata, the smaller labeled sample, the different construction of the transient and contaminant populations, and the narrower taxonomy used for the reported run.

\section{Discussion}\label{sec:discussion}
The benchmark results show that the neural-process backbone can model irregular, partially observed transient light curves in a classification setting. The joint reconstruction and classification objective provides a way to learn from sparse context, while contextual metadata contributes most when the alert history contains the least photometric information. 
%These findings support the inclusion of host information in broker-facing classification pipelines whenever it is available.
Both properties are relevant to broker deployment, though the dependence on metadata also makes the classifier sensitive to the host-population composition of the training set.

The current work also has clear limits, however. The primary benchmark remains simulated, and the taxonomy is intentionally collapsed from the full ELAsTiCC2 model inventory into broader physical families. The ELAsTiCC2 AGN-like bucket, for example, is 
%driven by CLAGN in the underlying simulation, not by a broad observationally heterogeneous AGN population. The resulting benchmark is a controlled Rubin-oriented test bed. 
populated by CLAGN alone, not by a broad observational AGN population. The resulting benchmark is a controlled Rubin-oriented test bed, not a complete representation of the real alert stream. The TDE simulations likewise remain approximations because observed TDEs show diversity in early bumps, color evolution, late-time plateaus, rebrightening, and decline rates that is difficult to capture with a compact template set.

\added{The ELAsTiCC2 class frequencies are controlled by simulation design, not expected on-sky alert-stream rates, and rare families are deliberately overrepresented to ensure training coverage. MALLORN instead begins from the class mixture of a spectroscopically classified ZTF nuclear-transient sample and is closer to an observed breakdown within that selected population. It is still shaped by the magnitude limit and by class-dependent follow-up and classification efficiencies, particularly below the Bright Transient Survey completeness regime \citep{2026RASTI...5ag019M}. Neither benchmark should therefore be treated as an exact forecast of the Rubin class mixture. Threshold-dependent metrics such as F1 will shift with the deployed class distribution. AUROC is less sensitive to prevalence, while average precision summarizes the purity--completeness trade-off for the rare target class. Both the operating threshold and realized precision will need to be recalibrated for Rubin alerts.}

The metadata gain also has limits. Host information improves classification, particularly under sparse context, but the classifier tends to favor source classes as they appear in the most common training-set environments. For TDE classification specifically, this means the model may perform less well for off-nuclear events or those arising in atypical host galaxies.

NAPTIME is also evaluated here as a closed-set classifier on a broad family-level taxonomy, so any source outside the modeled families is still assigned to the nearest in-scope label. In practical broker settings, that kind of classifier would need to sit alongside upstream filtering, abstention logic, or anomaly-detection machinery \citep{2025RASTI...4...54G}. The next substantive test is transfer to small real-survey samples, where the main question is how well a simulation-trained family classifier survives the domain shift from ELAsTiCC2 to observed alert streams. Recent transfer-learning results suggest that adaptation to real survey data can be achieved with much less labeled data than a full retraining would require \citep{2025MNRAS.542L.132G}, and early Rubin analyses already show what such classification pipelines look like on Data Preview 1 transient products \citep{Freeburn2025DP1}.

Another important limitation is interpretability. Feature-engineered systems such as those of \citet{bhardwaj2025photometric} and \citet{stein2024tdescore} make their decision logic comparatively transparent through explicit colors, timescales, or SHAP analyses. NAPTIME gains flexibility by learning directly from irregular photometry, but this comes at the cost of a less direct mapping to hand-crafted astrophysical features. %In the present analysis we partially bridge this gap through per-class confusion analysis, metadata ablations, and prefix sweeps, but richer interpretability analysis remains worth pursuing.
Direct numerical comparison to published ELAsTiCC, broker, ZTF, or MALLORN results is not straightforward because these studies generally use different taxonomies, train--test splits, metadata assumptions, time-window definitions, and operating metrics. Table~\ref{tab:external-context} provides numerical context for representative classifiers, while the matched metadata-aware and photometry-only NAPTIME runs remain the controlled comparison in this work.

The present analysis characterizes model behavior through per-class confusion structure, metadata ablations, and prefix sweeps, but it does not provide feature-level explanations comparable to those available in engineered-feature pipelines. NAPTIME also differs from purely discriminative classifiers because it is trained to predict withheld photometry as well as class labels. That predictive task is used primarily to regularize the representation learned from sparse light curves, as opposed to as a standalone forecasting product. Future applications could use the same predictive distribution for alert forecasting or follow-up planning, but the present evaluation focuses on classification and TDE retrieval.

\section{Conclusions}\label{sec:conclusion}
We have introduced NAPTIME, a neural-process-based framework for photometric transient classification under Rubin-like observing conditions. The model combines a ConvGNP-style backbone, a global latent path, an optional metadata branch, and a joint reconstruction-classification objective. Across MALLORN and ELAsTiCC2, it provides a unified way to model irregular multiband light curves while supporting both photometry-only and metadata-aware inference.

Our main empirical conclusions are as follows:
\begin{enumerate}
    \item On the 15-family ELAsTiCC2 benchmark, the metadata-aware model outperforms the matched photometry-only version by a clear margin, indicating that host and redshift context remain useful within the simulated population represented by the benchmark even when the photometric backbone is already strong.
    \item The benefit of metadata is largest in the early-context regime, when the alert history is still short and the light curve has not yet become strongly diagnostic on its own.
    \item The same family-level classifier remains effective for TDE one-vs-rest retrieval, so TDE-focused candidate ranking can be interpreted within the broader broker taxonomy instead of being separated into a different model family.
\end{enumerate}
Overall, these results support the use of neural processes as a practical probabilistic framework for Rubin-era alert classification under sparse, irregular, and progressively accumulating observational context.

\begin{acknowledgments}
N.M.E., K.D.F., and M.S. acknowledge support from NSF grant AST–2206164. N.M.E. acknowledges support from the Center for Astrophysical Surveys Graduate Fellowship. J.T.H acknowledges support provided by NASA through the NASA Hubble Fellowship grant HST-HF2-51577.001-A awarded by the Space Telescope Science Institute, which is operated by the Association of Universities for Research in Astronomy, Incorporated, under NASA contract NAS5-26555. S.C. acknowledges support received from the NASA FINESST program, Grant 80NSSC25K0312 and is grateful to the NSF-Simons AI Institute for the Sky for hosting the Rubin Alerts \& AI Hackathon, which helped facilitate this collaboration.
\end{acknowledgments}

\begin{rotatetable*}
\begin{deluxetable*}{llcccccc}
% \tabletypesize{\scriptsize}
\tablewidth{0pt}
\tablecaption{Representative published photometric-classification results, with NAPTIME results included for scale. These are not matched baselines; see Section~\ref{sec:discussion} for the comparability caveats.\label{tab:external-context}}
\tablehead{\colhead{Model} & \colhead{Task} & \colhead{F1} & \colhead{AUROC} & \colhead{AP} & \colhead{$P$} & \colhead{$R$} & \colhead{Ref.}}
\startdata
NAPTIME+MD & E2-Fam/TDE\tablenotemark{a} & 0.903 & 0.991 & 0.985 & 0.935 & 0.950 & This work \\
NAPTIME phot. & E2-Fam/TDE\tablenotemark{a} & 0.874 & 0.986 & 0.979 & 0.919 & 0.938 & This work \\
NAPTIME phot. & MAL/TDE\tablenotemark{b} & 0.693 & 0.958 & 0.730 & \nodata & \nodata & This work \\
ATAT & E20\tablenotemark{c} & $0.829\pm0.004$ & \nodata & \nodata & \nodata & \nodata & \citep{Cabrera2024} \\
ATAT-BHRF & E20\tablenotemark{c} & $0.794\pm0.001$ & \nodata & \nodata & \nodata & \nodata & \citep{Cabrera2024} \\
ORACLE & E-Hier\tablenotemark{d} & \nodata & \nodata & \nodata & 0.96--$>0.99$ & \nodata & \citep{2025ApJ...995....4S} \\
ORACLE & E19\tablenotemark{d} & \nodata & \nodata & \nodata & 0.83 & \nodata & \citep{2025ApJ...995....4S} \\
Fink/CATS & E-Broker\tablenotemark{e} & \nodata & \nodata & \nodata & $\geq0.93$ & \nodata & \citep{Fraga2024Fink} \\
Fink periodic & E-Per.\tablenotemark{e} & \nodata & \nodata & \nodata & $\geq0.98$ & $\geq0.99$ & \citep{Fraga2024Fink} \\
\texttt{tdescore} & ZTF-TDE\tablenotemark{f} & \nodata & \nodata & \nodata & 0.802 & 0.775 & \citep{stein2024tdescore} \\
ALeRCE-TDE & ZTF-TDE\tablenotemark{g} & \nodata & \nodata & \nodata & \nodata & 0.91 & \citep{PavezHerrera2025ALeRCETDE} \\
Bhardwaj & E2-TDE\tablenotemark{h} & \nodata & \nodata & \nodata & $\leq0.95$ & $\sim0.72$ & \citep{bhardwaj2025photometric} \\
FLEET-TDE & ZTF-TDE\tablenotemark{i} & \nodata & \nodata & \nodata & 0.30--0.50 & $\sim0.40$ & \citep{gomez2023fleet} \\
\enddata
\tablecomments{F1 and AUROC are macro-averaged where reported. AP is average precision. $P$ and $R$ denote precision and recall for the task named in the second column.}
\tablenotetext{a}{E2-Fam/TDE: ELAsTiCC2 family-level classification with TDE one-vs-rest metrics.}
\tablenotetext{b}{MAL/TDE: MALLORN multiclass classification with TDE one-vs-rest AP.}
\tablenotetext{c}{E20: 20-class ELAsTiCC classification.}
\tablenotetext{d}{E-Hier and E19: ORACLE hierarchical ELAsTiCC tasks; ORACLE reports macro-averaged precision, including 19-way precision at 1024 days after first detection.}
\tablenotetext{e}{E-Broker and E-Per.: Fink ELAsTiCC broker and periodic-class tasks.}
\tablenotetext{f}{ZTF-TDE: ZTF nuclear-transient TDE retrieval.}
\tablenotetext{g}{ZTF-TDE: ZTF broker taxonomy with TDE expansion.}
\tablenotetext{h}{E2-TDE: ELAsTiCC2 binary TDE retrieval.}
\tablenotetext{i}{ZTF-TDE: FLEET TDE selection at $p_{\rm TDE}>0.5$; precision is $\sim0.30$ using the first 20 days of photometry and $\sim0.50$ using 40 days, with similar completeness of $\sim0.40$.}
\end{deluxetable*}
\end{rotatetable*}

% \begin{contribution}
% This section is a placeholder for the author contribution statement and will be updated in a later revision.
% \end{contribution}

% \facilities{Rubin Observatory (simulated), ZTF (discussion only)}

% \software{astropy,
%           PyTorch,
%           scikit-learn,
%           matplotlib,
%           NumPy}

\appendix

\section{Reproducibility Details}\label{appendix:repro}
This appendix summarizes the dataset splits, model configuration, and runtime scale of the reported runs. All final runs use seed 42, a stratified validation fraction of 0.15, and full-context evaluation for the headline metrics. During training, random partial contexts are sampled with a context mask probability drawn from the configured range, while sequential prefix sweeps are used only for the early-context evaluation.

\subsection{Dataset Splits and Class Counts}
The uncapped ELAsTiCC2 experiments use all available objects in the family-level taxonomy in Table~\ref{tab:elasticc-taxonomy}, rather than the smaller class-capped subsets used for smoke tests and some exploratory runs. The resulting stratified split contains 1,313,370 training objects and 231,772 validation objects. Table~\ref{tab:appendix-elasticc-counts} lists the per-family counts.

\begin{deluxetable}{lrr}
\tablecaption{ELAsTiCC2 family-level train--validation counts used for the uncapped experiments.\label{tab:appendix-elasticc-counts}}
\tablehead{\colhead{Family} & \colhead{Train} & \colhead{Validation}}
\startdata
TDE & 80142 & 14143 \\
AGN-like & 92598 & 16341 \\
Ca-rich/fast & 9965 & 1759 \\
ILOT & 9059 & 1599 \\
Kilonova & 5374 & 948 \\
SLSN & 80248 & 14162 \\
PISN & 77211 & 13625 \\
SN~Ia & 215420 & 38015 \\
SN~II & 299834 & 52912 \\
SN~Ibc & 270629 & 47758 \\
M-dwarf flare & 2257 & 398 \\
Pulsator & 58835 & 10383 \\
Eclipsing binary & 80694 & 14240 \\
Dwarf nova & 9744 & 1720 \\
Microlensing & 21360 & 3769 \\
\enddata
\end{deluxetable}

The MALLORN experiment uses a six-class taxonomy derived from the labeled training release and computes TDE retrieval one-vs-rest from the multiclass posterior. The reduction groups SLSN-I and SLSN-II as SLSNe; \added{the normal, 91T-like, 91bg-like, peculiar, and Iax subtypes as SN~Ia}; SN~II, SN~IIn, and SN~IIP as SN~II; and SN~Ib, SN~Ic, SN~Ic-BL, and SN~IIb as SN~Ibc. \added{All labeled objects in the training release enter one of these six groups.} With the same seed and validation fraction, the split contains 2,586 training objects and 457 validation objects. Table~\ref{tab:appendix-mallorn-counts} lists the class counts.

The reported MALLORN scores are computed on this stratified validation split of the labeled training release rather than on the official challenge test set. We treat these values as an internal photometry-only benchmark and not as leaderboard-comparable scores. The validation split contains only 22 TDEs, so TDE-specific quantities such as one-vs-rest average precision have substantially larger sampling uncertainty than the ELAsTiCC2 metrics. The same validation split is also used for checkpoint selection and final reporting, which can introduce mild optimistic bias; this effect is expected to be negligible for the much larger ELAsTiCC2 validation set but is more relevant for MALLORN.

\begin{deluxetable}{llrr}
\tablecaption{MALLORN taxonomy reduction and train--validation counts used for the reported multiclass photometry-only experiment.\label{tab:appendix-mallorn-counts}}
\tablehead{\colhead{Class} & \colhead{Included MALLORN labels} & \colhead{Train} & \colhead{Validation}}
\startdata
TDE & TDE & 126 & 22 \\
AGN & AGN & 1518 & 268 \\
SLSN & SLSN-I, SLSN-II & 21 & 4 \\
SN~Ia & \begin{tabular}[t]{@{}l@{}}SN~Ia, \added{SN~Ia-91T-like, SN~Ia-91bg-like,}\\ \added{SN~Ia-pec,} SN~Iax\added{[02cx-like]}\end{tabular} & 700 & 124 \\
SN~II & SN~II, SN~IIn, SN~IIP & 178 & 32 \\
SN~Ibc & SN~Ib, SN~Ic, SN~Ic-BL, SN~IIb & 43 & 7 \\
\enddata
\end{deluxetable}

\subsection{Model and Optimization Configuration}
The final ELAsTiCC2 and MALLORN runs use the same neural-process backbone, with the metadata branch enabled only for the metadata-aware ELAsTiCC2 model. Table~\ref{tab:appendix-model-config} gives the checkpoint-verified architecture settings. The metadata-aware ELAsTiCC2 model has 5.65 million trainable parameters, the photometry-only ELAsTiCC2 model has 5.64 million, and the MALLORN model has 5.64 million.

\begin{deluxetable}{ll}
\tablecaption{NAPTIME architecture and optimization settings used for the final runs.\label{tab:appendix-model-config}}
\tablehead{\colhead{Quantity} & \colhead{Value}}
\startdata
Bands & 6 Rubin bands \\
Grid & 256 points on normalized time interval $[0,1]$ \\
Band embedding dimension & 8 \\
Fourier time-feature dimension & 8 \\
Point-feature dimension & 128 \\
Grid-feature dimension & 256 \\
Convolutional backbone & 8 layers, kernel size 5, hidden dimension 128 \\
Convolutional dropout & 0.1 \\
Gaussian set-convolution scales & 0.015, 0.030, 0.060 \\
Latent path & enabled, dimension 32, hidden dimension 128 \\
Classifier and decoder hidden dimensions & 128 \\
Metadata branch & 10 inputs, 64 hidden units, 32-dimensional embedding \\
Loss weights & $\lambda_{\rm recon}=1$, $\lambda_{\rm cls}=1$, $\beta_{\rm KL}=5\times10^{-4}$ \\
KL warm-up & 20 epochs \\
Optimizer & AdamW, learning rate $3\times10^{-4}$, weight decay $10^{-5}$ \\
Batch size & 32 \\
Checkpoint metric & macro F1 \\
\enddata
\end{deluxetable}

For ELAsTiCC2, the metadata vector consists of Milky Way reddening, host separation, dDLR, host stellar mass, host ellipticity, and host $gri zY$ magnitudes. Host photometric redshift is passed through the dedicated redshift input and is not duplicated in the auxiliary metadata vector. The photometry-only ELAsTiCC2 model keeps the same photometric backbone and redshift input but disables the metadata branch. MALLORN uses the same backbone with redshift enabled and metadata disabled.

\subsection{Runtime Scale}
Runtime measurements are hardware dependent, so they are best interpreted as scale estimates for the compute environment used in the final runs rather than as intrinsic model constants. Loading the full eager ELAsTiCC2 training set on the cluster took about one minute before optimization began. The metadata-aware ELAsTiCC2 checkpoint selected by macro F1 was reached at epoch 34, while the matched photometry-only checkpoint was reached at epoch 37. The full training jobs were run for 40 epochs, so any training-duration estimate should be read as an approximate wall-clock cost for this hardware and data-loading configuration.

Inference time is measured post hoc from the saved checkpoint by timing the model-evaluation pass over the full validation split, excluding the initial ELAsTiCC2 file loading and excluding CSV/JSON writing. On a local Apple M1 system using the PyTorch MPS backend, the metadata-aware ELAsTiCC2 model evaluated 231,772 validation objects in 661 s at batch size 32, corresponding to 2.85 ms per object.

\subsection{Evaluation Artifacts}
The reported metrics and figures were generated from the saved metadata-aware ELAsTiCC2, photometry-only ELAsTiCC2, and MALLORN checkpoints. ELAsTiCC2 full-context metrics, prefix sweeps, TDE operating-point curves, confusion matrices, and permutation-importance summaries all use the same seed-42 validation split described above. Figure rerendering used the saved prediction, threshold, and permutation artifacts rather than repeating model evaluation for style-only changes.

\bibliography{references}{}
\bibliographystyle{aasjournalv7}

\end{document}